\documentclass[pra,amsmath,amssymb,twocolumn]{revtex4}
\usepackage{color}
\usepackage{graphicx}
\usepackage{epsfig}
\usepackage{dcolumn}
\usepackage{bm}

\begin{document}
\title{Synchronization of micromasers}
\date{\today}
\author{C. Davis-Tilley and A. D. Armour}
\affiliation{Centre for the Mathematics and Theoretical Physics of Quantum Non-Equilibrium Systems and School of Physics and Astronomy, University of Nottingham, Nottingham NG7 2RD, UK}

\begin{abstract}
We investigate synchronization effects in quantum self-sustained oscillators theoretically using the micromaser as a model system. We use the probability distribution for the relative phase as a tool for quantifying the emergence of preferred phases when two micromasers are coupled together. Using perturbation theory, we show that the behavior of the phase distribution is strongly dependent on exactly how the oscillators are coupled. In the quantum regime where photon occupation numbers are low we find that although synchronization effects are rather weak, they are nevertheless significantly stronger than expected from a semiclassical description of the phase dynamics. We also compare the behavior of the phase distribution with the mutual information of the two oscillators and show that they can behave in rather different ways.
\end{abstract}
\maketitle
\section{Introduction}

Self-sustained oscillators do not have a preferred phase, but  when two or more of them are weakly coupled together a phase preference can emerge spontaneously, an effect known as synchronization. Self-sustained oscillators are ubiquitous in nature and synchronization effects have been widely studied across the physical and biological sciences\,\cite{pikovsky}. Synchronization has also been studied in quantum optical systems such as the laser, although generally focussing on regimes where approximate semiclassical descriptions work well\,\cite{cresser,fabiny}. In the last few years there has been considerable interest in studying the synchronization of oscillators and related systems\,\cite{zhirov,fazio13,ludwig,lee,zambrini,walter14,lee2,mendoza,walter,hush,holland,fazio,zhu,coupvdp,weiss,brandes,li,bruder,weiss2} close to threshold or at low excitation levels where semiclassical approaches break down and fully quantum mechanical calculations are required. Recent theoretical work has explored different ways of quantifying synchronization in quantum oscillators\,\cite{fazio13,lee,hush,fazio,li}, as well as investigating the connection between it and measures of correlation such as mutual information and entanglement\,\cite{fazio13,zambrini,lee2,fazio,brandes}. Detailed comparisons have also been made between the predictions of quantum models and those of related semiclassical or classical descriptions\,\cite{lee,walter}.

Studies of synchronization effects in the quantum regime have largely concentrated on the behavior of simple model systems such as van der Pol oscillators\,\cite{lee,walter14,lee2,walter,fazio,coupvdp,brandes,weiss2} (together with closely related models\,\cite{bruder}), though a number of other systems including atomic ensembles\,\cite{holland,zhu} and optomechanical oscillators\,\cite{fazio13,ludwig,weiss,walter} have also been investigated. In this article we investigate synchronization in a very different model system consisting of two weakly coupled micromasers.

The micromaser is a self-sustained oscillator consisting of a microwave cavity  driven by a steady flow of excited atoms which interact strongly with a particular cavity mode\,\cite{review,filipowicz,englert}. The micromaser was used to carry out a range of pioneering experiments in quantum optics\,\cite{review}. However, it has also become possible to engineer systems with similar behavior in the solid-state using, for example, superconducting\,\cite{armour,marthaler} or optomechanical\,\cite{nation} devices.

The micromaser makes a very interesting model system with which to explore synchronization effects in the quantum regime because it displays a very rich range of dynamical behaviors, including strongly non-classical features which go well beyond those found in simpler systems like the quantum van der Pol oscillator. Furthermore, an exact steady-state solution is available for the density operator of the micromaser\,\cite{filipowicz} and important dynamical properties such as the linewidth\,\cite{scully,quang,schieve,schieve2} have been studied extensively.

Using a probability distribution for the relative phase, we explore how a preference for a particular phase (or phases) emerges when two micromasers are weakly coupled together in different ways. We investigate the behavior of the phase distribution over a wide range of parameters ranging from a semiclassical regime where photon occupation numbers are large to a quantum regime where occupation numbers are small and the steady-state of the system can be strongly non-classical. We derive a simple Fokker-Planck equation for the phase distribution assuming large photon occupation numbers and compare it with numerical calculations using the full master equation of the system. We find that whilst the Fokker-Planck equation  provides a good description of the phase distribution in the semiclassical regime, it substantially underestimates the extent to which a preferred phase emerges in the quantum regime. We also compare the behavior of the phase distribution with the mutual information and entanglement of the micromasers and find that the behavior is somewhat different in each case.

This work is organized as follows. We introduce our model of the coupled micromaser system and briefly review the key properties of the uncoupled micromaser in Sec.\ \ref{sec:model}. Then we introduce the relative phase distribution in Sec.\ \ref{sec:relp}. We show how perturbation theory can be used to understand the behavior of the phase distribution in the weak coupling limit in Sec.\ \ref{sec:pert}. Then in Sec.\ \ref{sec:sc} we derive a simple analytic formula for the phase distribution in the semiclassical limit and compare it with numerical calculations. We examine the behavior of other measures of correlation between the micromasers in Sec.\ \ref{sec:mi}.
Finally, we summarize our findings and discuss possible directions for future work in Sec. \ref{sec:conc}.

\section{Micromaser Model}
\label{sec:model}

We consider a model system consisting of two micromasers, assumed for simplicity to be identical, that are coupled together weakly. As is the case for classical oscillators\,\cite{pikovsky,couplings2,couplings}, the behavior can be very sensitive to the form of the coupling as well as its strength.  We investigate two specific forms for the coupling involving either additional terms in the Hamiltonian of the system (coherent coupling) or additional dissipative terms in the master equation (dissipative coupling). The starting point for our model is the standard master equation description for the micromaser\,\cite{filipowicz,englert}, to which we shall simply add additional terms to describe the coupling.

The master equation for the density operator of the two micromasers, $\rho$, (in the interaction picture) takes the form
\begin{equation}
\dot{\rho}=\mathcal{L}_1[\rho]+\mathcal{L}_2[\rho]+\mathcal{L}_c[\rho] \label{eq:twomm}
\end{equation}
where  here $\mathcal{L}_1[\rho]$ and $\mathcal{L}_2[\rho]$ describe the uncoupled dynamics of the two micromasers and the interaction between them is given by $\mathcal{L}_c[\rho]$.

 The dynamics of each individual micromaser is controlled by a balance of interactions between the cavity and the flow of atoms which pass through it, and between the cavity and its electromagnetic environment which gives rise to losses. The atoms can be traced out of the master equation so that the dynamics of the system is captured by the terms\,\cite{filipowicz,englert}
\begin{eqnarray}
\mathcal{L}_j[\rho]&=&N\left[\cos(\phi\sqrt{a_ja_j^{\dagger}})\rho\cos(\phi\sqrt{a_ja_j^{\dagger}})\right.\\\nonumber
&&\left.+\frac{a_j^{\dagger}\sin(\phi\sqrt{a_ja_j^{\dagger}})}{\sqrt{a_ja_j^{\dagger}}}\rho\frac{\sin(\phi\sqrt{a_ja_j^{\dagger}})a_j}{\sqrt{a_ja_j^{\dagger}}}-\rho\right]\\ \nonumber
&&+\frac{1}{2}\left[2a_j\rho a_j^{\dagger}-a_j^{\dagger}a_j\rho-\rho a_j^{\dagger}a_j\right], \label{eq:mmme}
\end{eqnarray}
where $a_j$ is the lowering operator for a mode of the cavity $j$ ($j=1,2$), $N$ is the rate at which atoms pass through the cavity and $\phi$ is the Rabi angle which quantifies the strength of the atom-cavity interaction. Note that we have adopted units of time such that the cavity loss rate, $\gamma$, is unity and we have taken the zero-temperature limit for simplicity.

For coherent coupling  between the two micromasers the interaction is described by the Hamiltonian
\begin{equation}
H_c=\hbar\varepsilon(a_1a^{\dagger}_2+a^{\dagger}_1a_2),
\end{equation}
with $\varepsilon$ the coupling strength (scaled by $\gamma$), and hence in this case
 \begin{equation}
 \mathcal{L}_c^{\textrm{(coh)}}[\rho]=-\frac{i}{\hbar}[H_c,\rho].
 \end{equation}
 For dissipative coupling the master equation includes terms which describe an additional loss channel for the system whose properties depend on the state of both modes\,\cite{lee2,walter,coupvdp}
\begin{eqnarray}
\mathcal{L}_{c}^{\textrm{(diss)}}[\rho]&=&\varepsilon\left[(a_1-a_2)\rho(a_1^{\dagger}-a_2^{\dagger})\right.\nonumber\\
&&\left.-\frac{1}{2}\left\{(a_1^{\dagger}-a_2^{\dagger})(a_1-a_2),\rho\right\}\right]. \label{eq:dc}
\end{eqnarray}
In the following we will focus mainly on the regime where the coupling is very weak $\varepsilon\ll 1$.

\begin{figure}[t]
\centering{\includegraphics[width=8.0cm]{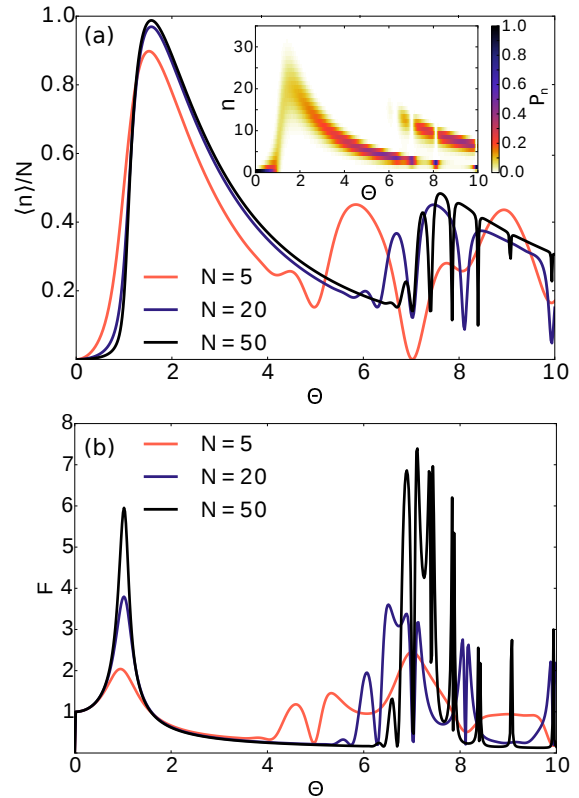}}

\caption{(Color Online) Steady-state occupation number $\langle n\rangle/N$ (a) and associated Fano factor, $F$, (b) for an uncoupled micromaser as a function of $\Theta=\phi N^{1/2}$ and $N$. The inset in (a) shows the full number distribution $P_n$ for the case where $N=20$.}
\label{fig:mm1}
\end{figure}

Before going on to look at how either coherent or dissipative coupling affects the system, we briefly review the most important properties of the uncoupled micromaser ($\varepsilon=0$).
The steady-state density operator for a micromaser is diagonal in the number state representation and the probability of finding the cavity in the $n$-th number state is\,\cite{filipowicz,englert}
\begin{eqnarray}
P_n=K\prod_{m=1}^{n}\frac{N\sin^2(\phi\sqrt{m})}{m},
\end{eqnarray}
where $K$ is a constant determined by normalization.

Although the density operator can be written in terms of an apparently simple formula, the state of the system changes dramatically as a function of the atomic flow rate, $N$, and the strength of the atom-cavity interaction, parameterized by the pump parameter: $\Theta=\phi N^{1/2}$.  Fig. \ref{fig:mm1} illustrates the behavior of the average occupation number $\langle n\rangle$ and the number fluctuations, measured by the Fano factor $F=(\langle n^2\rangle-\langle n\rangle^2)/\langle n\rangle$, as a function of $\Theta$ and $N$; the inset shows an example of the full number distribution $P_n$. The system has a threshold at $\Theta=1$, above which a limit-cycle emerges---manifesting in the $P_n$ distribution as a peak at non-zero $n$. After initially growing very rapidly in size, the limit-cycle then gets progressively smaller as $\Theta$ is increased (up until $\Theta\sim 4$). There is a strong peak in $F$ around threshold and it then drops below unity, a signature of number squeezing\,\cite{knightbook}.

For $\Theta > 4$ the behavior becomes more complicated and the micromaser moves between a range of different states. It undergoes dynamical transitions between limit-cycles with different average energies\,\cite{filipowicz} and can co-exist in a mixed state involving two limit-cycle states (seen as two peaks in the $P_n$ distribution). At certain specific values of $\phi$ such that $\sin(\phi\sqrt{m+1})=0$ with $m=0,1,2,...$ the system becomes \emph{trapped}:   $P_{(n>m)}=0$ because the matrix element that generates transitions between the $m$ and $m+1$ number states vanishes. These trapping states\,\cite{review} have a number state distribution that can be extremely sharply peaked. For example, at $\phi=\pi/\sqrt{2}$ ($m=1$ trapping state) one typically finds $P_1\gg P_0$ and as $N$ is increased the system gets closer and closer to being exactly in the $n=1$ number state.  

The extremely narrow $P_n$ distributions that the micromaser displays above threshold come close to reaching what might be thought of as the most quantum of limit-cycle states---pure number states. However, the micromaser's steady-state isn't always strongly non-classical. The average occupation number of the micromaser is roughly proportional to $N$ for fixed $\Theta$ and the non-classical features are strongest for either relatively small $N$ or larger values of $\Theta$. In contrast, when $\langle n\rangle\gg 1$ and $\langle n\rangle \gg \phi^2$ the dynamics of the number distribution, $P_n$, is well described by a Fokker-Planck equation in which $n$ is treated as a continuous variable\,\cite{filipowicz}, which corresponds to the semiclassical limit of the system.



\section{Relative phase distribution}
\label{sec:relp}
Phase distributions provide a convenient way of characterizing the emergence of a preferred relative phase in systems of coupled oscillators. For a single oscillator the quantum mechanical phase distribution is given by\,\cite{knightbook},
 \begin{equation}
  P(\varphi)=\frac{1}{2\pi}\langle \varphi| \rho|\varphi\rangle=\frac{1}{2\pi}\sum_{n,m=0}^{\infty}\langle n|\rho|m\rangle{\textrm{e}}^{i(m-n)\varphi},
  \end{equation}
  where $|\varphi\rangle=\sum_{n=0}^{\infty}{\textrm e}^{in\varphi}|n\rangle$ is an eigenstate of the Susskind-Glogower operator $\sum_{n=0}^{\infty}|n\rangle\langle n+1|$. This phase distribution, $P(\varphi)$, also emerges naturally from the Pegg-Barnett description of the phase operator\,\cite{pegg} or indeed when one seeks to define a distribution which is the canonical conjugate of the number distribution\,\cite{ulf}. In the steady-state only the diagonal components of the uncoupled micromaser density operator are non-zero in the number distribution, so we see immediately that there is no preferred phase and the phase distribution is simply uniform: $P(\varphi)=1/2\pi$.

  When two micromasers are coupled, either coherently or dissipatively, a preference emerges for certain values of the \emph{relative} phase, $\varphi_-=\varphi_{1}-\varphi_2$, but not the total phase $\varphi_+=\varphi_{1}+\varphi_2$. The relative phase distribution takes the form\,\cite{bp,ls,hush}
\begin{eqnarray}
P(\varphi_-)&=&\frac{1}{2\pi}\sum_{n,m=0}^{\infty}\sum_{k={\textrm{max}}(n,m)}^{\infty}{\textrm{e}}^{i\varphi_-(m-n)}\nonumber\\
&&\times\langle n,k-n|\rho|m,k-m\rangle,\label{eq:prelp}
\end{eqnarray}
which can also be rewritten explicitly as a Fourier series
\begin{equation}
P(\varphi_-)=\frac{1}{2\pi}+\frac{1}{\pi}{\textrm{Re}}\left[\sum_{p=1}^{\infty}{\textrm{e}}^{ip\varphi_-}\sum_{n=0,m=0}^{\infty}\rho^{(p)}_{n,m}\right], \label{eq:prelp2}
\end{equation}
where we adopt the notation $\rho_{n,m}^{(p)}=\langle n,m+p|\rho|n+p,m\rangle$.

It is also helpful to be able to characterize the emergence of a phase preference using a single number. Starting from the relative phase distribution, one can simply extract the size of the peak relative to the uniform distribution\,\cite{hush},
\begin{equation}
S= 2\pi\,\textrm{max}\left[P(\varphi_-)\right] - 1.
\end{equation}
This is something that we will make extensive use of here, though it should be noted that the choice is by no means unique\,\cite{weiss,bruder}.

\begin{figure}
\centering
{\includegraphics[width=8.0cm]{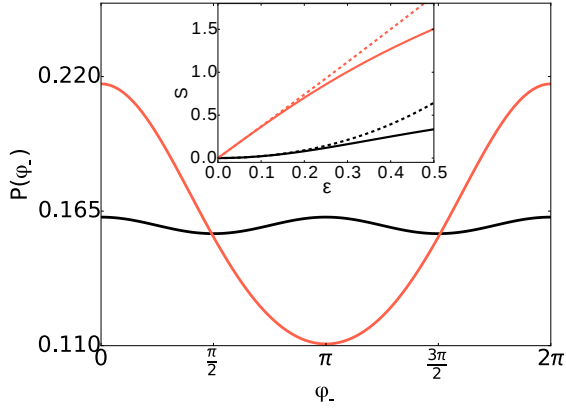}
}
\caption{(Color Online) Comparison of the relative phase probability distributions for coherent (black) and dissipative (red) coupling with $\varepsilon=0.1$, $N=5$ and  $\Theta=\phi N^{1/2}=2$. The inset shows the strengths of the peaks in the relative phase distributions, characterized by $S$, as a function of the coupling; the full lines are from full numerical calculations while the dashed lines are from the perturbation theory described in Sec.\ \ref{sec:pert}.}
\label{fig:phicomp}
\end{figure}

The relative phase distribution for the micromaser system is obtained by solving for the steady-state of the master equation \eqref{eq:twomm} using standard numerical methods\,\cite{qutip}. Coherent and dissipative couplings with the same strength give rise to markedly different behavior in the relative phase distribution, as is illustrated  in Fig.\ \ref{fig:phicomp}. Coherent coupling leads to much weaker phase locking than  dissipative coupling and generates a relative phase distribution which is $\pi$-periodic rather than $2\pi$-periodic. This matches the well-known differences in relative phase dynamics of reactively and dissipatively coupled classical oscillators which are usually understood by deriving approximate equations of motion for the relative phase of the oscillator assuming weak coupling\,\cite{pikovsky,couplings}.

\section{Perturbation Theory}
\label{sec:pert}
Perturbation theory provides a straightforward way of understanding the differences between the quantum mechanical relative phase distributions generated by coherent and dissipative coupling. We begin by considering the case of coherent coupling before moving on to dissipative coupling.

\subsection{Coherent Coupling}
 Writing the master equation in terms of the number state basis, we find

 \begin{eqnarray}
\dot{\rho}_{n,m}^{(p)}&=&-\left[\frac{\mu_n^{(p)}+\mu_m^{(p)}}{2}+c_n^{(p)}+d_n^{(p)}+c_m^{(p)}+d_m^{(p)}\right]\rho_{n,m}^{(p)} \nonumber \\ &&+c_{n-1}^{(p)}\rho_{n-1,m}^{(p)}+c_{m-1}^{(p)}\rho_{n,m-1}^{(p)}+d_{n+1}^{(p)}\rho_{n+1,m}^{(p)}\nonumber \\
&&+d_{m+1}^{(p)}\rho_{n,m+1}^{(p)}+\Delta^{(p)}_{n,m}, \label{eq:meqnnb}
\end{eqnarray}
where the terms arising from the coupling are given by
\begin{eqnarray}
\Delta^{(p)}_{n,m}&=&-i\varepsilon\left[\sqrt{n(m+p+1)}\rho^{(p+1)}_{n-1,m}\right.\nonumber\\
&&+\sqrt{(n+1)(m+p)}\rho^{(p-1)}_{n+1,m}\nonumber\\
&&-\sqrt{m(n+p+1)}\rho^{(p+1)}_{n,m-1}\nonumber\\
&&\left.-\sqrt{(n+p)(m+1)}\rho^{(p-1)}_{n,m+1}\right], \label{eq:delta0}
\end{eqnarray}
 and the coefficients which describe the uncoupled evolution are given by\,\cite{scully}
\begin{eqnarray}
\frac{1}{2}\mu_n^{(p)}&=&2N\sin^2\left[\frac{\phi}{2}\left(\sqrt{n+p+1}-\sqrt{n+1}\right)\right]\nonumber\\
&&+\left[n+\frac{p}{2}-\sqrt{n(n+p)}\right]\\
c_{n-1}^{(p)}&=&N\sin(\phi\sqrt{n})\sin(\phi\sqrt{n+p})\\
d_n^{(p)}&=&\sqrt{n(n+p)}.
\end{eqnarray}
In the steady-state $\dot{\rho}_{n,m}^{(p)}=0$, leading to a set of linear equations for the components ${\rho}_{n,m}^{(p)}$.  For $\varepsilon=0$ the sets of linear equations for different $p$ values are uncoupled; they are also homogeneous and have the solutions $\rho^{(p)}_{n,m}=0$, except in the diagonal case ($p=0$) where the components must also obey the normalization condition.

Working to first order in the coupling means that we replace the terms in Eq.\ \eqref{eq:delta0} by their unperturbed values which are all zero---apart from the diagonal ones, $\rho_{n,m}^{(0)}=P_nP_m$. Therefore only the equations for the $p=1$ components are affected (we only need consider the components with $p>0$ which appear in the expression for the relative phase distribution \eqref{eq:prelp2}) for which we find
\begin{equation}
\Delta^{(1)}_{n,m}=-i\varepsilon\sqrt{(n+1)(m+1)}\left(P_{n+1}P_{m}-P_{m+1}P_n\right). \label{eq:delta1}
\end{equation}
Thus at first order, the components $\rho^{(1)}_{n,m}$ are in general non-zero, and pure imaginary (since $P_{n(m)}$ are probabilities), whilst those with $p>1$ remain zero. However, since $\Delta^{(1)}_{n,m}=-\Delta^{(1)}_{m,n}$, there is a corresponding  relationship between the components: $\rho_{n,m}^{(1)}=-\rho_{m,n}^{(1)}$. This has important consequences for $P(\varphi_-)$: the sum of all the $p=1$ elements is zero and hence to first order $P(\varphi_-)=1/2\pi$.

Working to second order in the coupling, $\Delta^{(2)}_{n,m}$ is no longer zero (since $\rho_{n,m}^{(1)}$ is of order $\varepsilon$) and one finds that the components $\rho_{n,m}^{(2)}$ are all real and proportional to $\varepsilon^2$. In this case there is no cancelling of the terms in the sum and hence to second order the relative phase distribution takes the form
\begin{equation}
P(\varphi_-)=\frac{1}{2\pi}\left[1+\varepsilon^2 C_0\cos(2\varphi_-)\right],
\end{equation}
where $C_0$ is a constant which depends on the parameters of the uncoupled micromasers ($\phi$ and $N$). This of course is just what we see for the case of coherent coupling in Fig.\ \ref{fig:phicomp}.



\subsection{Dissipative coupling}
We now look at what happens for dissipative coupling described by Eq.\ \eqref{eq:dc}.
In this case we can simplify the master equation  by taking into account the fact that some of the terms in \eqref{eq:dc} simply act to increase the effective damping of the oscillators and can be absorbed into the terms which are already present for the uncoupled system by rescaling the parameters: $\tilde{N}=N/(1+\varepsilon)$, $\tilde{\varepsilon}=\varepsilon/(1+\varepsilon)$.

Working in the number basis, the master equation with dissipative coupling  takes the form given by \eqref{eq:meqnnb} (though  with $\tilde{\varepsilon}$ and $\tilde{N}$ replacing $\varepsilon$ and $N$) with
\begin{eqnarray}
\Delta^{(p)}_{n,m}&=&-\tilde{\varepsilon}\left[\sqrt{(n+1)(m+1)}\rho^{(p-1)}_{n+1,m+1}\right.\nonumber\\
&&+\sqrt{(n+p+1)(m+p+1)}\rho^{(p+1)}_{n,m}\nonumber\\
&&-\frac{1}{2}\sqrt{n(m+p+1)}\rho^{(p+1)}_{n-1,m}\nonumber\\
&&-\frac{1}{2}\sqrt{(n+p+1)m}\rho^{(p+1)}_{n,m-1}\nonumber\\
&&-\frac{1}{2}\sqrt{(n+1)(m+p)}\rho^{(p-1)}_{n+1,m}\nonumber\\
&&-\left.\frac{1}{2}\sqrt{(n+p)(m+1)}\rho^{(p-1)}_{n,m+1}\right]. \label{eq:dc2}
\end{eqnarray}
At first order in the coupling, only the $p=1$ term is non-zero,
\begin{eqnarray}
 \Delta^{(1)}_{n,m}&=& \frac{\tilde{\varepsilon}}{2}\sqrt{(n+1)(m+1)}\\
 &&\times\left[P_{n+1}P_m+P_nP_{m+1}-2P_{n+1}P_{m+1}\right]. \nonumber
\end{eqnarray}
This generates  non-zero components $\rho^{(1)}_{n,m}$ which are all real with $\rho^{(1)}_{n,m}=\rho^{(1)}_{m,n}$ so there is no cancellation when they are summed up.
Hence to lowest order in the coupling the relative phase distribution for dissipative coupling takes the form
\begin{equation}
P(\varphi_-)=\frac{1}{2\pi}\left[1+\varepsilon C_1\cos(\varphi_-)\right],
\end{equation}
where $C_1$ is a constant (that again depends on the parameters of the uncoupled micromasers) which matches the behavior in Fig.\ \ref{fig:phicomp}.

 The insights into the general form of the relative phase distribution provided by perturbation calculations are quite general, in the sense that the overall form of the relative phase distribution is entirely determined by the coupling terms (only the constants $C_0$ and $C_1$ depend on the details of the micromaser systems). However, perturbation theory is also very useful for exploring the behavior where the state space of the system becomes so large that a direct numerical solution becomes impracticable.

\section{Semiclassical Limit}
\label{sec:sc}

To understand the behavior in the semiclassical limit we start from the expression for the relative phase distribution in the form of Eq.\ \eqref{eq:prelp2} and focus on the case of dissipative coupling, which is the simplest. The equation of motion for the relative phase is given by
\begin{equation}\label{eq:prelp3}
\dot{P}(\varphi_-,t)=\frac{1}{\pi}{\textrm{Re}}\left[\sum_{p=1}^{\infty}{\textrm{e}}^{ip\varphi_-}\sum_{n,m=0}^{\infty}\dot{\rho}_{n,m}^{(p)}\right].
\end{equation}
Using the master equation \eqref{eq:meqnnb}, with the dissipative coupling terms \eqref{eq:dc2}, we find that\,\cite{schieve}
\begin{eqnarray}
& \sum_{p=1}^{\infty}{\textrm{e}}^{ip\varphi_-}\sum_{n,m=0}^{\infty}\dot{\rho}_{n,m}^{(p)}= \sum_{n,m=0}^{\infty}\rho_{nm}^{(0)}f_0(\tilde{\varepsilon})&\label{eq:sum}\\
&+\sum_{p=1}^{\infty}{\textrm{e}}^{ip\varphi_-}\sum_{n,m=0}^{\infty}\left[f_1(\tilde{\varepsilon})-\frac{1}{2}\left(\mu_n^{(p)}+\mu_m^{(p)}\right)\right]{\rho}_{n,m}^{(p)}& , \nonumber
\end{eqnarray}
where
\begin{equation}
f_0(\tilde{\varepsilon})=\frac{1}{2}\tilde{\varepsilon}{{\textrm{e}}^{i\varphi_-}}\left[\sqrt{n(m+1)}+\sqrt{m(n+1)}-2\sqrt{mn}\right]
\end{equation}
and
\begin{eqnarray}
&f_1(\tilde{\varepsilon})=\frac{1}{2}\tilde{\varepsilon}\left[{\textrm{e}}^{i\varphi_-}\left(\sqrt{n(m+p+1)}+\sqrt{m(n+p+1)}\right.\right.&\nonumber\\
&\left.\left.-2\sqrt{nm}\right)-{\textrm{e}}^{-i\varphi_-}\left(2\sqrt{(m+p)(n+p)}\right.\right.&\nonumber\\
&\left.\left.-\sqrt{(m+p)(n+1)}-\sqrt{(n+p)(m+1)}\right)\right].&
\end{eqnarray}

So far the analysis has been exact. We now introduce some approximations to simplify things\,\cite{scully,schieve}. We will make use of the fact that the average occupation is large in the semiclassical regime so that $\langle n\rangle\gg 1$ and $\langle n\rangle\gg \phi^2$. We will also make two additional assumptions: that the coupling is weak $\tilde{\varepsilon}\ll 1$ and that the number distribution is strongly peaked about $\langle n\rangle$, this will only be the case when the system is above threshold with $\Theta>1$, but not so large that more than one peak appears in the distribution [see the inset in Fig.\ \ref{fig:mm1}]. These last two assumptions have nothing to do with semiclassicality (indeed a strongly peaked number distribution can be highly non-classical), but rather they are closely related to those usually made in classical analyses of synchronization\,\cite{pikovsky,stratonovich} which rely on the existence of a well defined (single) limit-cycle state.

Our assumptions mean that we need only consider  the components $\rho_{n,m}^{(p)}$ for which $n,m\gg p$.
  We proceed by expanding the coefficients in \eqref{eq:sum} treating $p/n$, $p/m$, $1/n$ and $1/m$ as small quantities and keeping the lowest order (non-zero) contributions in each case so that we have
\begin{eqnarray}
f_0(\tilde{\varepsilon})&\simeq &\frac{1}{4}\tilde{\varepsilon}{{\textrm{e}}^{i\varphi_-}}\left(\frac{n+m}{\sqrt{nm}}\right)\\
f_1(\tilde{\varepsilon})&\simeq&\frac{1}{4}\tilde{\varepsilon}\left[{\textrm{e}}^{i\varphi_-}(p+1)+{\textrm{e}}^{-i\varphi_-}(p-1)\right]\nonumber\\
&&\times \left(\frac{n+m}{\sqrt{nm}}\right)
\end{eqnarray}
and
\begin{eqnarray}
\mu_n^{(p)}&\simeq &\left[4\tilde{N}\sin^2\left(\frac{p\phi}{4\sqrt{n+1}}\right)+\frac{p^2}{4 n}\right] \nonumber \\
&\simeq&\left[\frac{\tilde{N}p^2\phi^2}{4(n+1)}+\frac{p^2}{4 n}\right].
\end{eqnarray}
In the last line we also expanded the sine assuming $(p\phi)^2/n \ll 1$.

Finally, we make use of the assumption that the distributions are strongly peaked about a common average\,\cite{scully,schieve} and simply replace $n$ and $m$ (together with $n+1$ and $m+1$) with $\langle n\rangle$ so that now Eq.\ \eqref{eq:sum} takes the simplified form
\begin{eqnarray}
\sum_{p=1}^{\infty}{\textrm{e}}^{ip\varphi_-}\sum_{n,m=0}^{\infty}\dot{\rho}_{n,m}^{(p)}&=&\frac{1}{2}\tilde{\varepsilon}{{\textrm{e}}^{i\varphi_-}} -\sum_{p=1}^{\infty}{\textrm{e}}^{ip\varphi_-}\sum_{n,m=0}^{\infty}{\rho}_{n,m}^{(p)}\nonumber\\
&&\times\left[p^2\tilde{\Delta}-\tilde{\varepsilon}(\cos\varphi_-+ip\sin\varphi_-)\right],\nonumber\\ \label{eq:sumb}
\end{eqnarray}
where we have defined
\begin{equation}
\tilde{\Delta}=\frac{\tilde{N}\phi^2+1}{4\langle n\rangle}.
\end{equation}
The parameter $\tilde{\Delta}$ matches a simple approximate expression for the micromaser  linewidth\,\cite{scully,quang,schieve,schieve2} which is valid in the semiclassical regime.

Combining Eq.\,\eqref{eq:sumb} with its complex conjugate leads to  a Fokker-Planck equation for the relative phase distribution:
\begin{equation}
\dot{P}(\varphi_-)=\left(\frac{\partial}{\partial\varphi_-}\tilde{\varepsilon}\sin\varphi_-+\tilde{\Delta}\frac{\partial^2}{\partial\varphi_-^2}\right)P(\varphi_-).
\end{equation}
The corresponding steady-state distribution is\,\cite{stratonovich}
\begin{equation}\label{eq:FP}
P(\varphi_-)=\frac{1}{2\pi I_0(\tilde{\varepsilon}/\tilde{\Delta})}{\textrm e}^{\tilde{\varepsilon}\cos\varphi_-/\tilde{\Delta}}.
\end{equation}

The Fokker-Planck equation for the phase distribution of the coupled micromasers is exactly what we would expect for coupled classical oscillators in the presence of noise\,\cite{pikovsky,stratonovich}: it describes a competition between phase diffusion and the effects of the coupling which tends to drive the system towards a particular relative phase. However, the origin of the noise which drives the diffusion is nevertheless ultimately quantum mechanical rather than classical and hence it makes sense to see \eqref{eq:FP} as a semiclassical equation in this context. It is worth noting that a Fokker-Planck equation with the same form emerges in the analysis of coupled lasers far above threshold\,\cite{fabiny}.

\begin{figure}[t]
\centering
{\includegraphics[width=8.0cm]{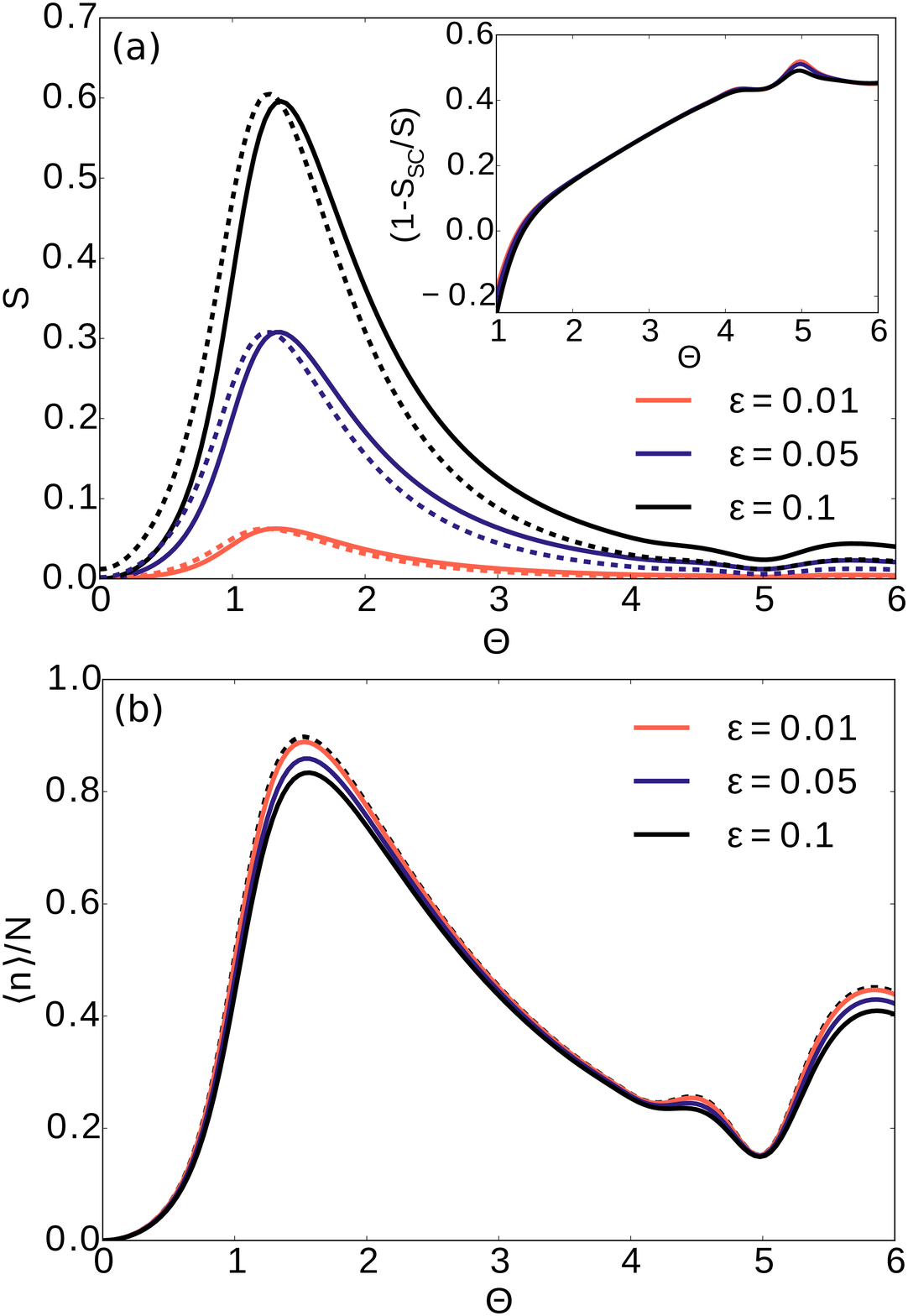}}

\caption{(Color Online)(a) Strength of the relative phase locking, measured by $S=2\pi[P(\varphi_-)_{\textrm{max}}]-1$, as a function of $\Theta=\phi N^{1/2}$ for dissipative coupling with $N=5$ and $\varepsilon=0.01$, $0.05$ and $0.1$. In each case the results of full numerical calculations are shown as full lines and the results from a semiclassical calculation using the Fokker-Planck equation are dashed lines. The inset shows the relative difference between the quantum and semiclassical calculations for the same parameters. [Note that the small peak around $\Theta=5$ corresponds to the $n=1$ trapping state] (b) Behavior of the average photon number for the same parameters compared with the uncoupled case, $\varepsilon=0$ (dashed line).}
\label{fig:sc1}
\end{figure}

Now that we have obtained an expression for the phase distribution in the semiclassical limit we can look in detail at when and how its predictions differ from the full (quantum) dynamics predicted by the master equation.
The phase distribution obtained using the semiclassical approximation (\ref{eq:FP}) is compared with the results from a full numerical solution of the master equation for a relatively small pumping rate, $N=5$, in Fig.\ \ref{fig:sc1}. The first thing to note is that the standard expectation of classical synchronization theory is fulfilled: for the weak couplings used here a rather strong change in the phase distribution is combined with a relatively weak change in the average occupation number of the system. Furthermore, the semiclassical phase distribution does a reasonable job of describing the strength of the peak in the phase distribution for $\Theta$ values that are not far above threshold. However, the semiclassical calculation systematically underestimates the strength of the phase locking in the quantum regime where $\langle n\rangle\sim 1$ (i.e.\ $\Theta > 3$) as shown in the inset to Fig.\ \ref{fig:sc1}a. Although the phase locking is pretty weak for larger values of $\Theta$, it is nevertheless about twice as strong as predicted by the simple semiclassical calculation.

\begin{figure}[t]
\centering
{\includegraphics[width=8.0cm]{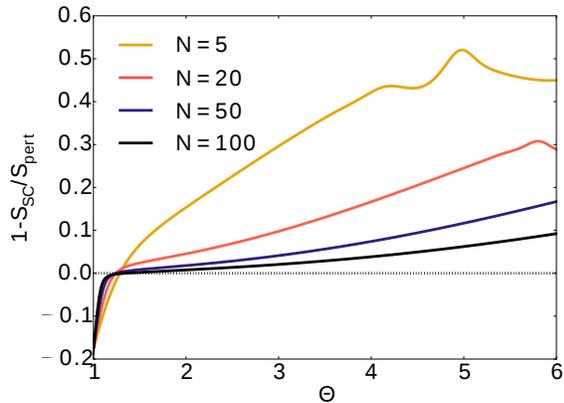}
}
\caption{(Color Online) Relative difference between the quantum (perturbation theory) and semiclassical calculations of the strength of the phase locking, $S$, for different values of $N$ with $\varepsilon=0.0001$.}
\label{fig:sc2}
\end{figure}

Although a full numerical solution of the master equation becomes very difficult for larger $N$ values we can use perturbation theory to calculate the relative phase distribution provided that we choose a small enough value for the coupling (since it turns out that the range of coupling strengths over which the second order perturbation calculation is a good description varies with $N$). Fig. \ref{fig:sc2} compares the value of $S$ obtained using the perturbation and semiclassical calculations  for a range of $N$ values. It shows clearly that the semiclassical solution \eqref{eq:FP} provides an increasingly accurate description of the strength of the phase locking as $N$ is increased, just as one would expect.

\section{Mutual information and entanglement}
\label{sec:mi}
We now turn to look at how other measures of correlations, mutual information and entanglement, behave as a function of the type and strength of the coupling between micromasers.  Recent work on coupled quantum van der Pol oscillators has suggested that there may be an intimate connection between the emergence of synchronization in coupled quantum oscillators and the behavior of the mutual information\,\cite{zambrini,fazio,brandes} and it has even been suggested that mutual information could serve as an order parameter for synchronization\,\cite{fazio}.

The mutual information, $\mathcal{I}$, of the coupled micromasers is defined as
\begin{equation}
\mathcal{I}=\mathcal{S}(\rho_1)+\mathcal{S}(\rho_2)-\mathcal{S}(\rho),
\end{equation}
where $\rho$ is the full density operator, $\rho_{1(2)}$, is the reduced density operator of micromaser $1(2)$ and $\mathcal{S}$ is the von Neumann entropy, $\mathcal{S}(\rho)=-{\textrm {Tr}}[\rho \ln\rho]$.

Perturbation theory tells us immediately that the mutual information of micromasers will grow at least quadratically with the strength of the coupling, for both dissipative and coherent coupling.  To see this, we can write the density operator $\rho$, its eigenvalues $\lambda_j$, and eigenkets $|j\rangle$, as expansions in\,\cite{footnote} $\varepsilon$
\begin{eqnarray}
\rho&=&\rho^{(0)}+\varepsilon\rho^{(1)}+\dots\\
\lambda_j&=&\lambda_j^{0}+\varepsilon \lambda_j^{1}+\dots\\
|j\rangle&=&|j^{(0)}\rangle+\varepsilon |j^{(1)}\rangle +\dots
\end{eqnarray}
To first order, the von Neumann entropy is
\begin{equation}
\mathcal{S}(\rho)=\mathcal{S}(\rho^{(0)})-\varepsilon\sum_j \lambda_j^{(1)}\left(1+\ln \lambda_j^{(0)}\right),
\end{equation}
with $\lambda_j^{(1)}=\langle j^{(0)}|\rho^{(1)}|j^{(0)}\rangle$, the first order correction to the eigenvalues.
 The density operators for uncoupled micromasers are diagonal in the number state basis,
  \begin{equation}
  \rho^{(0)}=\sum_{n,m}P_nP_m|n,m\rangle\langle n,m |,
  \end{equation}
  and as we have seen in Sec.\ \ref{sec:pert} the first order correction terms which form $\rho^{(1)}$ are
  all off-diagonal in the number state basis; consequently the first order corrections to the eigenvalues are all zero. Hence the first order contributions to $\mathcal{I}$ vanish for both coherent and dissipative couplings.

\begin{figure}[t]
\centering
{\includegraphics[width=8.0cm]{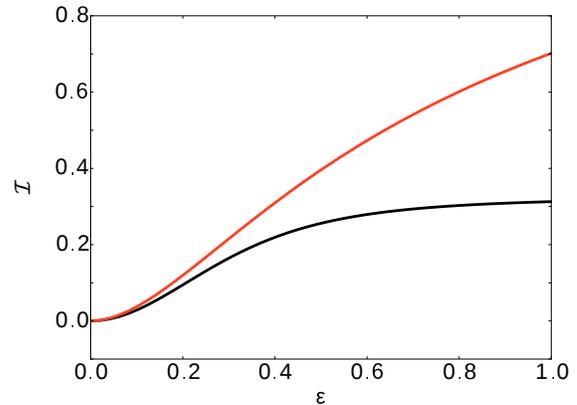}
}\caption{(Color Online) Mutual information as a function of the coupling strength for coherent (black) and dissipative (red) couplings. Here $N=5$ and $\phi N^{1/2}=2$.}
\label{fig:mi}
\end{figure}

Fig. \ref{fig:mi} compares the behavior of the  mutual information for dissipative and coherent couplings. Not only does $\mathcal{I}$  grow quadratically with $\varepsilon$ in both cases, but the magnitudes are very similar. This is in sharp contrast to the behavior of the relative phase distributions where the dissipative coupling leads to much stronger features in $P(\varphi_-)$ than the coherent coupling [see Fig.\ \ref{fig:phicomp}], with a linear rather than a quadratic dependence on $\varepsilon$.

These results suggest that the mutual information and the relative phase distribution characterize rather different aspects of the state of the coupled system. Whilst it is certainly to be expected that different measures of correlation between the micromasers will increase with the strength of the coupling, there is no reason why they should increase at the same rate.

\begin{figure}[t]
\centering
{\includegraphics[width=8.0cm]{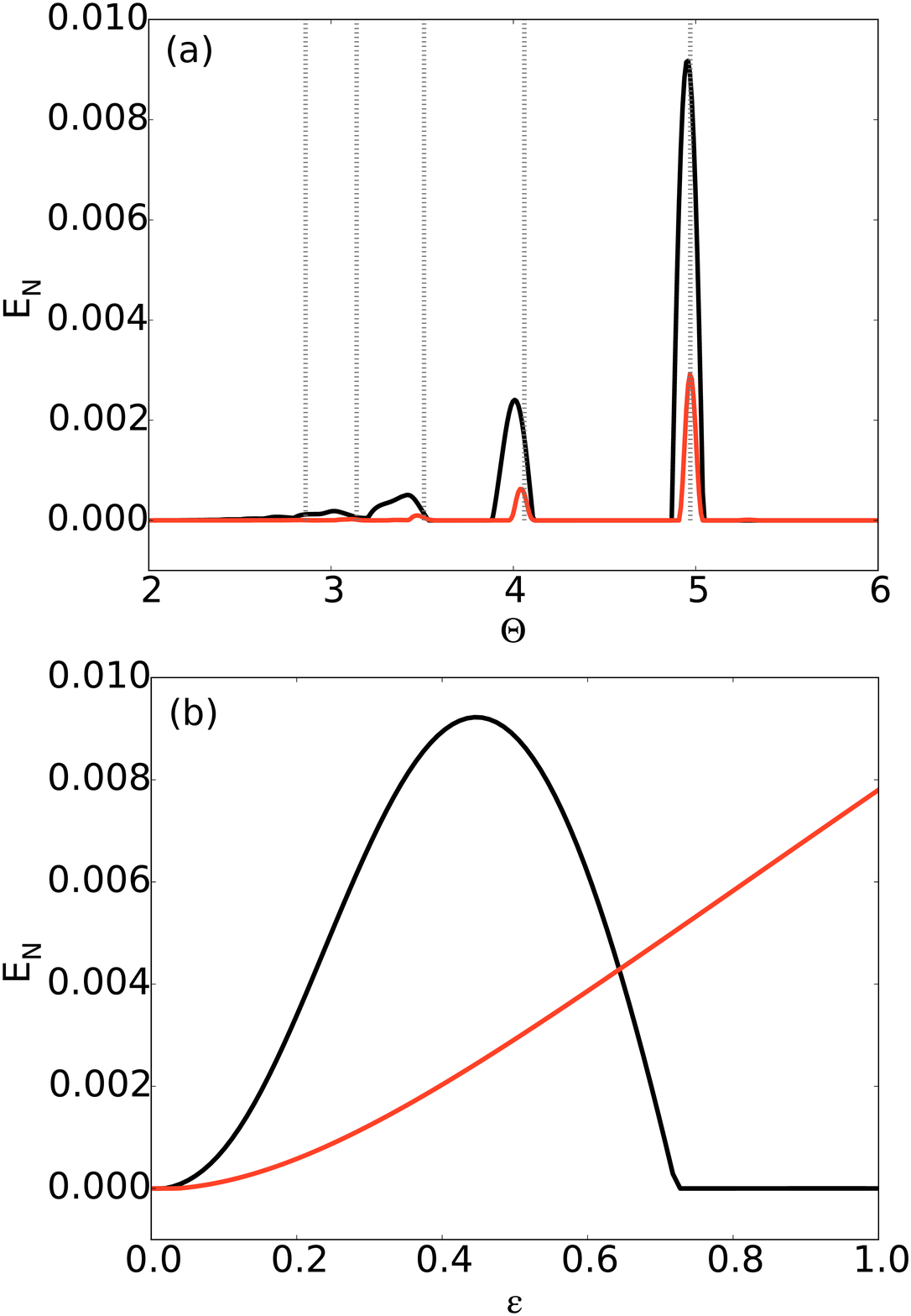}
}\caption{(Color Online) Entanglement for the coherent (black) and dissipative (red) couplings for $N=5$. (a) Logarithmic negativity for $\varepsilon=0.5$ over a range of $\Theta=\phi N^{1/2}$. The dotted lines indicate the locations of the trapping states ($n=1,2,3,4,5$). (b) Logarithmic negativity as a function of the coupling strength varied around the trapping state at $\Theta=4.97$.}
\label{fig:en}
\end{figure}

Finally, we comment briefly on the extent to which entanglement is generated between coupled micromasers. We use the logarithmic negativity of the system to measure the entanglement\,\cite{vidal},
\begin{equation}
E_N(\rho) = \textrm{log}_2 [ 2N(\rho) +1 ],
\end{equation}
where the negativity is $N(\rho) = \frac{1}{2} \sum_i \left( |\lambda_i| - \lambda_i \right)$, with $\lambda_i$ the eigenvalues of  $\rho^{T_A}$, the partial transpose of the density operator of the coupled micromaser system. Figure~\ref{fig:en} shows how the entanglement behaves  for both types of coupling.

We see in Fig.~\ref{fig:en}(a) that very little entanglement is in fact generated in this system. For both forms of coupling there is no entanglement except, interestingly, for the values of $\Theta$ that correspond to trapping states which occur at integer values, $n$, such that $\sin(\phi\sqrt{n+1})=0$. Figure \ref{fig:en}(b) shows how the entanglement at the $n=1$ trapping state ($\Theta=4.97$ for $N=5$) changes with the coupling strength $\varepsilon$, and we see that there is very different behavior depending on how the micromasers are coupled. However, for both cases the entanglement remains very weak, even in the strongly quantum regime of very low photon numbers.

\section{Conclusions and Discussion}
\label{sec:conc}
We have investigated synchronization effects in coupled micromasers using the relative phase distribution as a tool to quantify the emergence of preferred relative phases. We used perturbation theory to show that dissipative coupling between the micromasers leads to a $2\pi$-periodic relative phase distribution whose peak grows linearly with the coupling. In contrast, for coherent coupling the phase distribution is $\pi$-periodic and there is a quadratic dependence on the coupling.

The relative phase distribution seems to be a very useful tool for describing synchronization effects in the quantum regime. We were able to derive a Fokker-Planck equation to describe its dynamics in the semiclassical regime, where photon numbers are large, leading to a steady-state distribution which depended on just the coupling strength and the linewidth of the uncoupled micromaser. This derivation also showed quite clearly that the phase dynamics would be much more complicated in the quantum regime where photon occupation numbers are low. Indeed, comparisons of numerical calculations using the full master equation showed that the Fokker-Planck equation substantially underestimated the strength of the features which emerge in the relative phase distribution in the quantum regime. Interestingly, a very similar underestimate of synchronization effects was obtained using a semiclassical model for the case of two van der Pol oscillators\,\cite{lee}. In our case, it seems that low photon occupation number is the key factor that leads to differences between quantum and semiclassical predictions.

We also investigated the behavior of the mutual information of coupled micromasers and found that the behavior was rather different to that of the relative phase distribution. This is perhaps not surprising as the relative phase distribution depends on a very specific combination of a sub-set of the elements of the full density matrix of the system. Whilst all forms of correlation can be expected to increase with coupling, at least initially, it seems unlikely that other measures of correlation between the oscillators will increase in precisely the same way as the relative phase distribution.

Our work could serve as a starting point for a number of future studies.
For example, it would be interesting to explore synchronization in the bistable regime where the micromasers exist in a mixed state consisting of limit-cycles with two different amplitudes. Another possibility would be to  explore synchronization effects in systems with more than two micromasers. The perturbation approach that we used here could prove a useful tool in analysing systems with a handful of coupled oscillators where a numerical solution of the full master equation already becomes very challenging because of the potentially very large state space involved.
Finally, it would be interesting to investigate in detail the range of couplings which could be achieved in practice with micromasers, as well as  solid-state analogs, and the best way to measure features in their relative phase distribution.


\begin{thebibliography}{99}
\bibitem{pikovsky} A. Pikovsky, M. Rosenblum, and J. Kurths, {\it{Synchronization: A Universal Concept in Nonlinear Sciences}}, Cambridge Nonlinear
Science Series (Cambridge University Press, Cambridge, UK, 2003).
\bibitem{cresser} J. D. Cresser, W. H. Louisell, P. Meystre, W. Schleich and M. O. Scully, Phys. Rev. A {\bf{25}}, 2214 (1982).
\bibitem{fabiny} L. Fabiny, P. Colet, R. Roy and D. Lenstra, Phys. Rev. A {\bf{47}}, 4287 (1993).

\bibitem{zhirov} O.V. Zhirov and D.L. Shepelyansky Phys. Rev. Lett. 100 014101 (2008).
\bibitem{fazio13}  A. Mari, A. Farace, N. Didier, V. Giovannetti and R. Fazio, Phys. Rev. Lett. {\bf 111}, 103605 (2013).
\bibitem{ludwig} M. Ludwig and F. Marquardt
Phys. Rev. Lett. {\bf{111}}, 073603 (2013).
\bibitem{lee} T. E. Lee and H. R. Sadeghpour, Phys. Rev. Lett. {\bf 111}, 234101 (2013).
\bibitem{zambrini} G. Manzano, F. Galve, G. L. Giorgi, E. Hernández-Garcia, and R. Zambrini, Sci. Rep. {\bf{3}}, 1439 (2013).
\bibitem{walter14} S. Walter, A. Nunnenkamp and C. Bruder, Phys. Rev. Lett. {\bf 112}, 094102 (2014).

\bibitem{lee2} T. E. Lee, C.-K. Chan and S. Wang, Phys. Rev. E {\bf 89}, 022913 (2014).
\bibitem{mendoza} I. Hermoso de Mendoza, L. A. Pach\'{o}n, J. G\'{o}mez-Garde\~{n}es, and D. Zueco
Phys. Rev. E {\bf{90}}, 052904 (2014).
\bibitem{walter} S. Walter, A. Nunnenkamp and C. Bruder, Ann. Phys. {\bf 527}, 131 (2015).
\bibitem{holland} M. Xu, D. A. Tieri, E. C. Fine, J. K. Thompson, and M. J. Holland
Phys. Rev. Lett. {\bf 113}, 154101 (2014).
\bibitem{hush} M. R. Hush, W. Li, S. Genway, I. Lesanovsky and A. D. Armour, Phys. Rev. A {\bf{91}}, 061401 (2015).

\bibitem{fazio} V. Ameri, M. Eghbali-Arani, A. Mari, A. Farace, F. Kheirandish, V. Giovannetti and R. Fazio, Phys. Rev. A. {\bf 91} 012301 (2015).
\bibitem{zhu} B. Zhu, J. Schachenmayer, M. Xu, F. Herrera, J. G. Restrepo, M. J. Holland and A. M. Rey, New J. Phys. {\bf 17}, 083063 (2015).
\bibitem{coupvdp} L. Morgan and H. Hinrichsen, J. Stat. Mech., P09009 (2015).
\bibitem{weiss} T. Weiss, A. Kronwald and F. Marquardt, New J. Phys. {\bf 18}, 013043 (2016).
\bibitem{brandes}V. M. Bastidas, I. Omelchenko, A. Zakharova, E. Sch\"{o}ll and T. Brandes, Phys. Rev. E {\bf{92}}, 062924 (2015).
\bibitem{li} W. Li, C. Li and H. Song, Phys. Rev. E {\bf 93}, 062221 (2016).
\bibitem{weiss2}	T. Weiss, S. Walter and F. Marquardt, e-print arXiv:1608.03550.

\bibitem{bruder} N. L\"{o}rch, E. Amitai, A. Nunnenkamp and C. Bruder, Phys. Rev. Lett. {\bf 117}, 073601 (2016).


\bibitem{review} H. Walther, B. T. H. Varcoe, B.-G. Englert and T. Becker, Rep. Prog. Phys. {\bf{69}}, 1325 (2006).
\bibitem{filipowicz} P. Filipowicz, J. Javanainen and P. Meystre, Phys. Rev. A {\bf{34}}, 3077 (1986).
\bibitem{englert} B. G. Englert, e-print arXiv:quant-ph/0203052; B. G. Englert
and G. Morigi, Lect. Not. Phys. {f\bf{611}}, 55 (2002).

\bibitem{scully} M. O. Scully, H. Walther, G. S. Agarwal, T. Quang and W. Schleich, Phys. Rev. A {\bf{44}}, 5992 (1991).
\bibitem{quang} T. Quang, G. S. Agarwal, J. Bergou, M. O. Scully, H. Walther, K. Vogel and W. P. Schleich, Phys. Rev. A {\bf 48}, 803 (1993).
\bibitem{schieve} W. C. Schieve and R. R. McGowan, Phys. Rev. A {\bf{48}}, 2315 (1993).
\bibitem{schieve2} R. R. McGowan and W. C. Schieve, Phys. Rev. A {\bf{55}}, 3813 (1997).


\bibitem{armour} D. A. Rodrigues, J. Imbers, and A. D. Armour, Phys. Rev. Lett. {\bf 98}, 067204 (2007).
\bibitem{marthaler} M. Marthaler, J. Lepp\"akangas, and J. H. Cole, Phys. Rev. B {\bf 83}, 180505 (2011).
\bibitem{nation} P. D. Nation, Phys. Rev. A {\bf{88}} 053828 (2013).

\bibitem{couplings2} D. G. Aronson, G. B. Ermentrout and N. Kopell, Physica D {\bf 41}, 403 (1990).
\bibitem{couplings} A. Kuznetsov, N. Stankevich and L. Turukina, Physica D: Nonlinear Phenomena {\bf 238}, 1203 (2009).

\bibitem{knightbook}  C. C. Gerry and P. L. Knight, \emph{Introductory Quantum Optics}, (Cambridge University Press, Cambridge, UK, 2004).
\bibitem{pegg} D. T. Pegg and S. M. Barnett, Phys. Rev. A {\bf{39}}, 1665 (1989).
\bibitem{ulf} U. Leonhardt, \emph{Measuring the Quantum State of Light} (Cambridge University Press, Cambridge, UK, 1997).

\bibitem{bp} S. M. Barnett and D. T. Pegg, Phys. Rev. A {\bf 42}, 6713 (1990).
\bibitem{ls} A. Luis and L. L. S\'{a}nchez-Soto, Phys. Rev. A {\bf 53} 495 (1996).


\bibitem{qutip} J. R. Johansson, P. D. Nation and F. Nori, Comp. Phys. Comm. {\bf 183}, 1760 (2012); J. R. Johansson, P. D. Nation, and F. Nori, Comp. Phys. Comm. {\bf 184}, 1234 (2013).

\bibitem{stratonovich} R. L. Stratonovich, \emph{Topics in the Theory of Random Noise, Vol. II} (Gordon Breach, New York, 1967).
\bibitem{footnote} For dissipative coupling the expansion is in terms of $\tilde{\varepsilon}$ and the terms in Eq.\ \ref{eq:dc2} form the perturbation.
\bibitem{vidal} G. Vidal and R.F. Werner, Phys. Rev. A {\bf 65}, 032314 (2002).

\end{thebibliography}
\end{document}